# High-accuracy disposable micro-optical anti-counterfeiting labels based on single-molecule quantum coherence

Shuangping Han[1,2], Kai Song[1,2], Pengyu Zan[1,2], Changzhi Yu[1,2], Ao Li[1,2], Haitao Zhou[3], Chengbing Qin[1,4,*], Liantuan Xiao[1,2,4,*]

[1.] *College of Physics and Optoelectronics Engineering, Taiyuan University of Technology, Taiyuan, Shanxi 030024, China;*

[2.] *Key Laboratory of Advanced Transducers and Intelligent Control System of Ministry of Education, Taiyuan University of Technology, Taiyuan 030024, China;*

[3.] *Department of Nuclear Medicine, The First Hospital of Shanxi Medical University, Collaborative Innovation Center of Molecular Imaging Precision Medical, Shanxi Medical University, Taiyuan 030001, China;*

[4.] *State Key Laboratory of Quantum Optics and Quantum Optics Devices, Institute of Laser Spectroscopy, Shanxi University, Taiyuan, Shanxi 030006, China.*

\* *chbqin@sxu.edu.cn (C. Q.); xiaoliantuan@tyut.edu.cn (L. X.)*

**Abstract:**

In this work we introduce an innovative approach to single-molecule quantum coherence (SMQC)-based disposable micro-optical anti-counterfeiting labels. This method facilitates the editing and reading of anti-counterfeiting with single molecules used as the anti-counterfeiting information labels. The label is meticulously crafted through inkjet printing technology, while its authentication is achieved via frequency domain imaging. Through a validation process including experimental demonstration, numerical simulation, and neural network analysis, we demonstrate the feasibility of this approach, further validate the integrity of the miniature anti-counterfeiting information storage, and verify the signal extraction accuracy with the recognition accuracy of the labels is consistently above 99.995%. The combination of SMQC-based disposable micro-optical anti-counterfeiting technology is expected to enable more precise preparation of single-molecule-array chips, thus providing a crucial foundation for the advancement of high-tech and smart manufacturing industries.

# 1. Introduction:

Anti-counterfeiting technology has gradually become critical to ensure product safety and traceability [1, 2]. Among many anti-counterfeiting technologies, disposable anti-counterfeiting labels have attracted much attention in recent years due to their unique characteristics. The most significant feature is the non-replicability after use, thus effectively avoiding the possibility of secondary utilization and providing higher credibility for product authenticity verification. Therefore, researchers in the field of disposable anti-counterfeiting labels have invested a lot of effort [3-9]. The commonly used disposable anti-counterfeiting label technologies include paper material technology, radio frequency identification technology (RFID), quick response code (QR code), and bar code technology. Paper material technology uses special paper with high barrier properties to prevent labels from being torn off and then pasted again. Meanwhile, through the special coating technology, the label surface is resistant to various imitation methods [6, 10, 11]. RFID technology uses wireless radio frequency to carry out non-contact two-way data communication, so as to achieve the purpose of identifying targets and data exchange [5, 12, 13]. QR codes and bar codes carry more information through high-density geometric structures [2-4, 14-16].

As the coding process of the above methods is deterministic and easy to replicate, resulting in low security, the physical unclonable function (PUF) provides a new idea for anti-counterfeiting strategies [1, 17-19]. PUF-based anti-counterfeiting labels have been the focus of related research attention, as they are capable of detecting counterfeit goods with remarkable sensitivity and accuracy. Additionally, they possess favorable disposability characteristics and have a wide range of potential applications in various fields, including high-value goods protection and enhancing security measures for pharmaceutical and food products. Wang's group proposes an optical PUF-based anti-counterfeiting label from perovskite micro-laser arrays, where randomness is introduced through vapor-induced microcavity deformation [20]. Lin *et al.* successfully synthesized $Mn^{2+}/Er^{3+}$-co-doped $Cs_2AgInCl_6$ double perovskites via a chemical method and investigated the dual emission properties of the material [21]. They observed that the material exhibited a pronounced photochromism phenomenon in the presence of ultraviolet irradiation and delved deeply into the underlying mechanisms of dual emission and photochromism, concluding that the material has significant potential for anti-counterfeiting applications. Jianyu *et al.* proposed an efficient and stable PUF certification based on structural color-based PUF labels (SCPLs) prepared by injection casting colloidal crystals [22]. Utilizing the responsiveness of SCPLs, the PUF features can be switched between invisible and visible states to hide PUF information. Xiao *et al.* prepared photochromic PUF labels by randomly dispersing highly crystalline carbon nitride particles in polyvinyl acetate [23]. The color of the PVAc/HCCN films originating from photochromism is used as the first layer of safety, and the pattern of the randomly distributed HCCN particles within the photochromic area is used as the second layer of security. However, the broad spectrum and environmental sensitivity of optical materials make the signal difficult to distinguish, and the fidelity needs to be improved. Moreover, the cluttered background information can easily overwhelm the coded signal, which poses a challenge to the preparation, storage, and use of anti-counterfeiting labels. However, the materials employed in the aforementioned methods are costly to prepare [24], susceptible to physical damage [22], and exhibit broad-spectrum and environmentally sensitive characteristics [25], which collectively result in the generation of signals that are challenging to differentiate with fidelity and security.

In this work, we propose a disposable micro-optical anti-counterfeiting label based on the manipulation of the single-molecule quantum coherence (SMQC). The inherent coherence characteristics of single molecules are used as anti-counterfeiting information carriers. Based on theoretical calculation, the direct relationship between the emission intensity of dye molecules and the phase difference of the femtosecond pulses is pointed out. The modulating factor and the frequency domain imaging are extracted by discrete Fourier transform (DFT). The anti-counterfeiting information is constructed through the spatial distribution of single molecules by inkjet printing technology. By adding quantum dots (QDs) as interference signals on single molecules, we proved the scheme's anti-noise ability and information-hiding ability. In addition, the feasibility of quantum coherent information as an anti-counterfeiting label is verified by numerical simulation. Furthermore, the light-quenching property of single molecules was employed as an optical PUF to fabricate disposable security labels. The correlation between the number of single molecules and the accuracy of information decoding was also validated.

## 2. Principle of anti-counterfeiting based on SMQC

In our previous work, it has been shown that single molecules have good coherence properties, and that quantum coherence visibility can be extracted from the time evolution of the population probability $\rho_{ee}$ of the excited state of a single molecule by frequency-domain modulation.

Considering the presence of realistic dephasing, $\rho_{ee}$ can be expressed as:

$$\rho_{ee} = \frac{1}{2} \cdot \sin^2\theta \cdot (1 + V \cdot \cos\Delta\varphi) \tag{1},$$

where $V$ is the quantum coherence visibility, $\Delta\varphi$ is the relative phase. When the pulse width is considerably smaller than the time scale of single-molecule dynamics, the quantum coherence visibility can be expressed as follows: $V = e^{-\Delta t/T_2^*}$, in which $T_2^*$ is the dephasing time. Equation (1) indicates that the PL intensity of molecules is proportional to electron population probability $\rho_{ee}$, the quantum coherence properties can be explored by performing their PL intensity. However, the PL intensity of molecules is usually influenced by many factors, including chemical and physical stability (photon blinking) [26], transition dipole moment

[27], and energy transfer within the surrounding environment [28]. The periodic modulation of the relative phase Δφ effectively eliminates these effects, and the quantum coherence information in the PL can be rapidly obtained by DFT, as illustrated in Fig. 1c. The modulation intensity $f(\omega)$ at frequency $\omega/2\pi$ can be obtained by performing DFT on the arrival time of each photon:

$$f(\omega) = \left| \sum_{n=1}^{N} e^{-i \cdot t_n \cdot \omega} \right| \quad (2).$$

Where $N$ is the total number of photons. When $\omega t \gg 2\pi$, the relationship between the single-molecule modulation intensity and the total photon number and quantum coherence visibility is:

$$f(\omega) = \frac{1}{2} V \cdot N \quad (3).$$

Through this method, quantum coherence information and the corresponding frequency-domain imaging can be quickly obtained. As illustrated in Fig. 1d, the respective frequency spectrum of the dye molecule (squaraine-derived rotaxane (SR)) and the background were obtained through modulation measurements. A particularly straightforward outcome is that at the modulation frequency (1 kHz), the modulation intensity of the single molecule is markedly greater than that of the background. Figure 1e shows the time-domain (wide-field) PL imaging of SR prepared by spin coating, while Fig. 1f depicts the frequency-domain imaging calculated by DFT. It can be found that the imaging contrast in the frequency-domain is significantly improved compared with time-domain imaging, which is consistent with previous reports [29, 30]. Our recent works show that this method can achieve more than 300-fold improvement in imaging contrast under the premise of strong background, showing great potential in suppressing background noise [27, 31].

From the analysis of SMQC, it is clear that coherence visibility is crucial for distinguishing the signal from the background. And the quantum coherence properties cannot be read by PL imaging or other conventional methods. Based on this, we propose an anti-counterfeiting method that utilizes samples with different coherence visibility to achieve time-domain concealment of target information and precise extraction of information in the frequency domain. Specifically, inkjet printing technology is used to arrange single molecules according to the spatial distribution of the target information, and a second layer with lower coherence visibility and a disordered distribution is applied to mask the target information. Due to the difference in coherence visibility between the two layers, the concealed information in the time domain can be successfully extracted from the frequency-domain image. In the following, we will use single molecules and quantum dots as examples to analyze the causes of their coherence visibility differences and theoretically validate the feasibility of this anti-counterfeiting approach.

The dipole moment is one of the primary driving forces behind dephasing, as its presence facilitates the coupling of a quantum system to the surrounding environment [32]. The strength of this coupling is determined by both the magnitude and orientation of the dipole moment. Assuming that the ambient noise acts on the system dipole moment through random electric field fluctuations $\overline{E}(t)$, the auto-correlation function of the environment is [33, 34]:

$$\langle \overline{E}_i(t)\overline{E}_i(t+\tau)\rangle = \int J_i(\omega)e^{i\omega\tau}d\omega \tag{4},$$

where $\tau$ is the time delay, $J_i(\omega)$ is the noise spectral density, describing the coupling strength of the environment in direction i. $T_2^*$ is determined by the value of $J_i(\omega)$ at zero frequency.

Given the independence of the contributions from each direction, the total dephasing rate is a superposition of the three directions:

$$\frac{1}{T_2^*} = \frac{1}{T_{2,x}} + \frac{1}{T_{2,y}} + \frac{1}{T_{2,z}} \propto J_x(0) + J_y(0) + J_z(0) \tag{5},$$

where $T_{2,x}$, $T_{2,y}$, $T_{2,z}$ are the dephasing times in the x, y, and z directions, respectively.

The direction of the dipole moment of a single molecule is primarily determined by the distribution of electrons within the molecule. This is attributable to the rigid structure of the molecule, which results in a dipole moment in space that typically exhibits a single definitive direction, assuming that along the z-direction, then [34, 35]:

$$\frac{1}{T_2^{SM}} \propto J_z(0) \tag{6}.$$

If the ambient noise is white noise independent in all directions with intensity $S_{SM}$, then the dephasing time of single molecules can be written as:

$$T_2^{SM} = J_z(\omega) = \frac{1}{S_{SM}} \tag{7}.$$

QDs are typically quasi-spherical, ellipsoidal, or possess certain lattice symmetries, allowing the dipole moment to exist independently along three spatial directions. This structural property introduces multi-channel environmental noise effects, leading to an accelerated dephase rate. If the ambient noise is white noise independent in all directions, i.e., $J_i(\omega) = S_{QD}$ ($i = x,y,z$), then the total dephasing rate and time can be expressed as:

$$\frac{1}{T_2^{QD}} = 3S_{QD} \tag{8},$$

$$T_2^{QD} = \frac{1}{3S_{QD}} \tag{9}.$$

Moreover, the PL of QDs is relatively complex, involving multiple physical processes such as bright and dark excitons, biexcitons, and gray states [36, 37]. These processes cause the coherent information of QDs to become obscured by other incoherent behaviors. In summary, QDs exhibit short dephasing times, typically in the picosecond range [34, 38]. In contrast, single molecules have significantly longer dephasing times, generally in the nanosecond range.

The dephasing time $T_2^*$ characterizes the timescale over which a quantum system loses coherence, reflecting the rate at which environmental interactions induce phase randomization. During dephasing, quantum coherence visibility decays exponentially with time. A shorter $T_2^*$ indicates faster environment-induced coherence loss, leading to a rapid decline in coherence visibility and making it challenging to preserve quantum properties. Conversely, a longer $T_2^*$ allows the system to maintain coherence for a more extended period, enhancing the observability of coherence visibility in experiments. This property enables efficient anti-counterfeiting applications through careful material selection. In this paper, we selected SR molecules as anti-counterfeiting information carriers and CdTeSe/ZnS QDs (CdTeSe/ZnS QDs@800nm, 800QDs) as interference information. The contrast of frequency-domain imaging is adjusted by using the natural difference between the two in dephasing time.

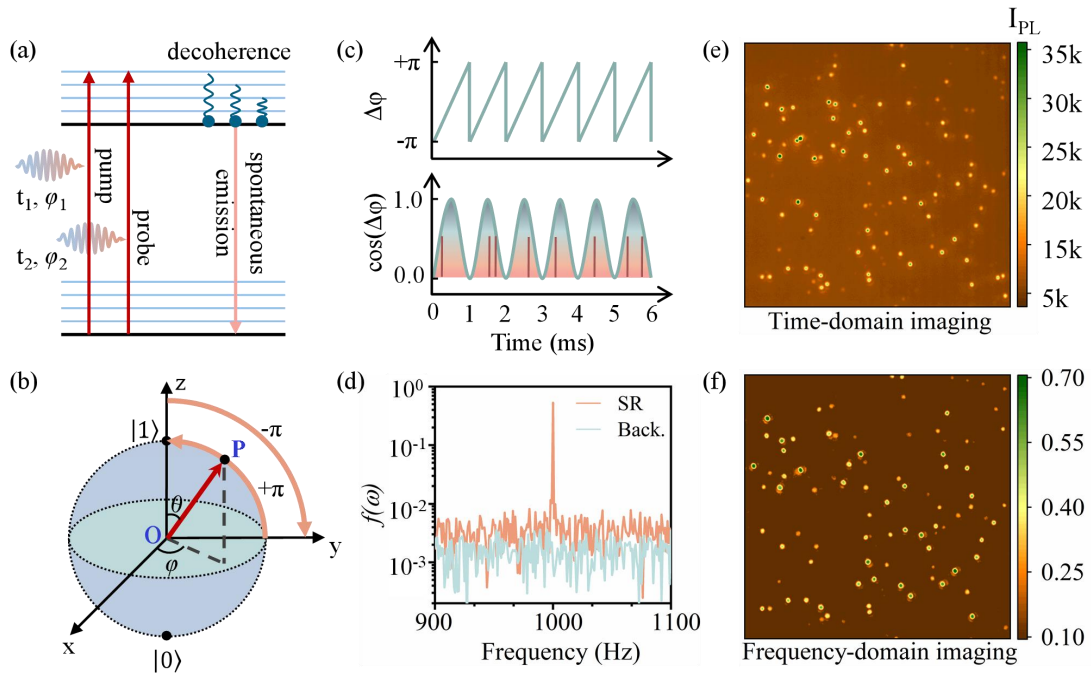

**Fig. 1. Theoretical model and frequency-domain imaging.** (a) Schematic diagram of the energy level for a single molecule. (b) Bloch sphere, the two poles represent the ground state and the excited state. In the absence of dephasing, the ±π pulse can make the electron transition between two eigenstates. (c) The periodic triangular wave signal used to modulate the laser phase and the cosine signal after loading on the electro-optic modulator (EOM). (d) Normalized Frequency spectrum of SR molecules and background measured by modulation measurements. (e, f) Time-domain imaging (Wide-field imaging) and frequency-domain imaging of spin-coated SR, respectively.

## 3. Neural network training for enhancing recognition accuracy

Based on this theory and phenomenon, single molecule labels can be prepared on the substrate by inkjet printing, and the writing and reading of anti-counterfeiting information can be realized by acquiring molecular coherence information through wide-field fluorescence imaging technology. In order to enhance the precision of recognition, we investigated the correlation between the number of single molecules present within the coding region and the degree of recognition accuracy, utilizing convolutional neural networks. The neural network used in this work is EfficientNet V2 (Details are in ESI), as shown in Figs. 2a-c. EfficientNet V2 further improves the resource utilization of EfficientNet V1 neural network by introducing Fused-MBConv and Inverted Residual Blocks [39]. By optimizing the depth, width, and resolution of the network, a better balance between computational efficiency and performance is achieved. An incremental learning strategy is adopted to enable the model to better adapt to complex training data and improve generalization ability and accuracy. Related conclusions can be found in the section entitled "Accuracy of information recognition".

## 4. Disposable micro-optical anti-counterfeiting labels based on SMQC

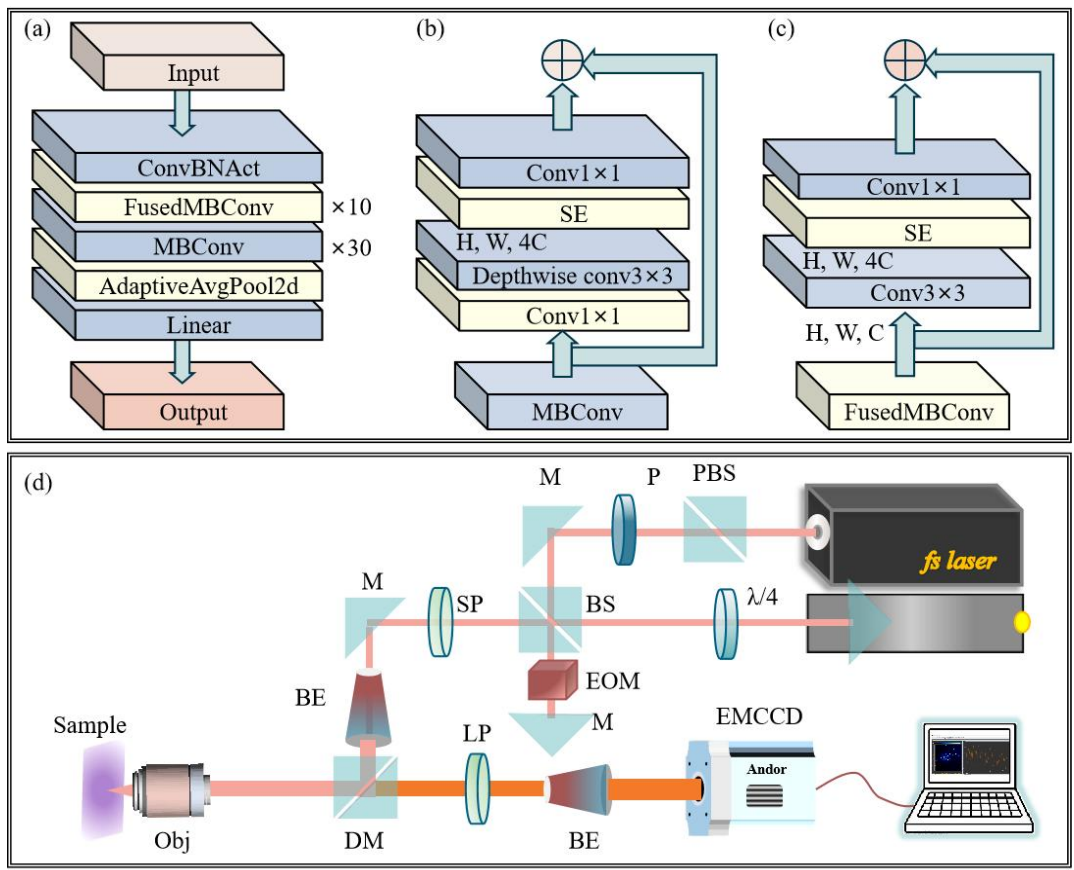

**Fig. 2. (a-c) Schematic diagram of EfficientNet V2. (d) Schematical diagram of the experimental setup.** M: mirror; P: polarizer; PBS: polarization beam splitter; BS: beam splitter; SP: short-pass filter; LP: long-pass filter; BE: beam expander; DM: dichroic mirror; Obj: objective lens.

To detect the quantum coherence dynamics of single molecules, we set up the experimental system, as shown in Fig. 2d. The ultrafast pulse used in the experiment is generated by a femtosecond laser from Toptica (FemtoFiber Pro, the pulse width is 389 fs). The laser is separated into two beams by a beam splitter (BS), one of them is used for phase modulation through an electro-optic modulator (EOM), and the other beam deflects the laser to vertical polarization through a quarter-wave plate (QWP). After that, the two beams are combined to produce pulse pairs with cross-polarization and adjustable phase. In this case, the intensity modulation caused by self-interference can be eliminated. By applying a periodic sawtooth wave voltage on the EOM, the relative phase between the pulse pairs is periodically modulated between $-\pi$ and $\pi$. The ultrafast pulse pairs are reflected into the objective lens through a dichroic mirror (DM) and then focused on the sample. Photons from the single molecule and background are collected by the objective lens and finally detected by EMCCD (Andor). The SMQC is obtained by performing DFT on the photon intensity of each pixel.

The results of employing this system to evaluate and illustrate the advantages of SMQC imaging in regard to imaging contrast are presented in Figs. 1e and 1f. The apparent light-quenching properties of single molecules present a significant challenge to the advancement of imaging technology. Nevertheless, this property is employed in this study to achieve the non-repeatability of the written label information, which serves as an effective carrier for the preparation of disposable optical security labels. Fig. 3b illustrates the time-domain imaging of SR molecules in the same measurement region at two distinct time points: t = 0.1 s and t = 1 s. It depicts a notable decline in the number of molecules, from a count exceeding 150 to a value below 20. This decline contributes to a swift reduction in recognition accuracy.

To verify the feasibility of the proposed scheme, we designed a disposable anti-counterfeiting method. The accurate information (English capital letter "H") is written by inkjet printing technology. However, the visualized letter can easily be detected by PL imaging. To achieve the anti-counterfeiting effect, we spin-coated a layer of 800QDs on the surface of the prepared single-molecule sample, and the specific structure is shown in Fig. 3a. The single molecules were separated from the QDs by polymethylmethacrylate (PMMA). Fig. 3c displays the result of time-domain imaging of the anti-counterfeiting label with the letter "H". The hidden written information is obscured by the luminescence of randomly distributed QDs. Subsequently, DFT calculations are performed on the collected photons, with the modulation factor employed to map the frequency-domain imaging. As shown in Fig. 3d, the frequency-domain imaging preserves the written information "H" while effectively removing the interfering information (800 QDs).

Based on this experiment, we proposed an anti-counterfeiting technology: encoding the position information of molecules through inkjet printing technology; adding another layer of luminescent material to hide the signal; and extracting the labels by SMQC. Our demonstration experiment verified the effectiveness of SMQC in shielding interference information. It is evident that the read accuracy of the information written by inkjet printing is directly dependent on the number of molecules. In order to enhance the precision of information transfer, we employ numerical simulations in conjunction with neural network training to assess the impact of molecule number on recognition accuracy in the following section.

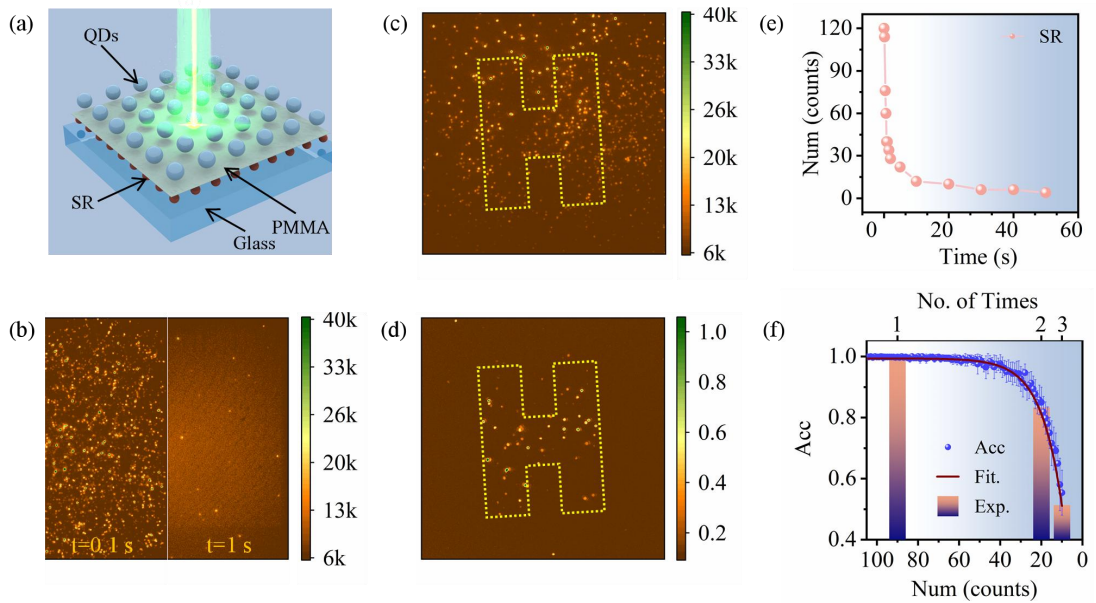

**Fig. 3. SMQC-based disposable anti-counterfeiting labels.** (a) Structure illustration for anti-counterfeiting labels. Substrate: glass; first layer: SR molecules; second layer: polymethylmethacrylate (PMMA); third layer: QDs. (b) Time-domain and frequency-domain imaging of spin-coated SR. (c) Time-domain imaging of the prepared SR label with a capital letter "H", where it can be found that the presence of interfering signals (800 QDs) makes the information of the letter "H" unobservable. (d) Frequency-domain imaging of Fig. 3c, the label "H" can be clearly extracted from the image carrying the interference information. (e) The decay trajectory of the number of SR molecules on glass with illumination time. (f) The recognition accuracy varies as a function of the molecular number and the number of measurements. The number of molecules decreases to ~20 by the third measurement, with recognition accuracy as low as 55%.

## 5. Accuracy of information recognition

The results of the demonstration experiment verified the effectiveness of the miniature optical anti-counterfeiting labels based on SMQC. Nevertheless, the quantity of effective input information, *i.e.*, the number of single molecules, represents a pivotal parameter in the signal recognition process. From Fig. 3e and related studies, it is known that the relationship between the number of surviving molecules and the laser power, and the duration of the laser

irradiation follows the following equation [40]:

$$N(t) = N_0 e^{-k_{bleach} \cdot t} = N_0 e^{-\alpha p \cdot t} \qquad (10),$$

Where $N(t)$ and $N_0$ are the number of molecules at time $t$ and the initial moment, respectively; $k_{bleach}$ is the rate of photobleaching, which is usually linear with the excitation power $p$, i.e., $k_{bleach}=\alpha p$, in which $\alpha$ is a constant of proportionality.

It is evident that the fewer the molecules that comprise the anti-counterfeiting information, the lower the recognition accuracy. However, this result is only at the level of qualitative description. Therefore, exploring the quantitative relationship between the number of single molecules and the recognition accuracy is helpful in providing a basis for the production of single-molecule anti-counterfeiting labels.

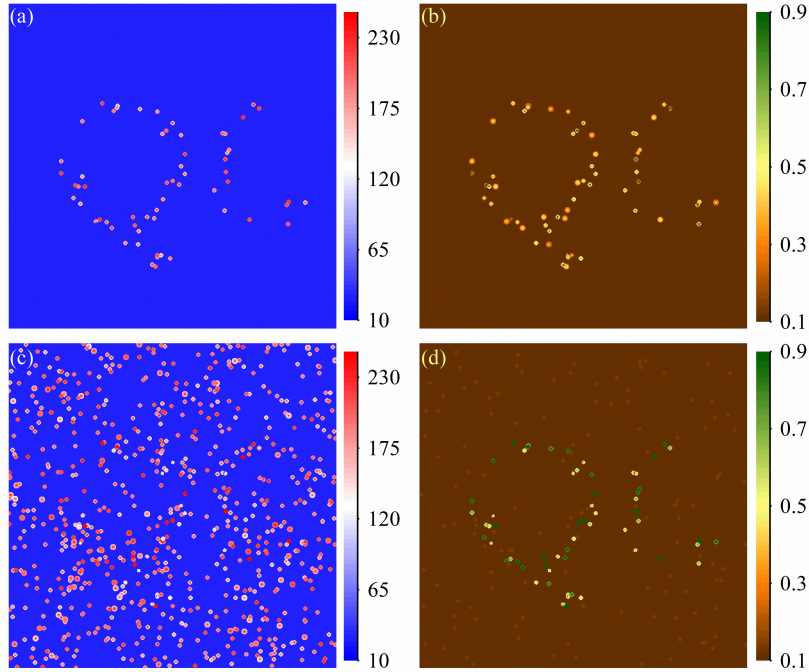

**Fig. 4. Numerical simulation of anti-counterfeiting labels based on SMQC.** (a, b) Time-domain imaging and frequency-domain imaging of the anti-counterfeiting label "QC". (c, d) Time-domain imaging and frequency-domain imaging under interference background.

In order to achieve the above purpose, we first reproduce the above demonstration experiment. Initially, we outline the character area (20825 pixels) on the background image (512×512 pixels, corresponding to 80×80 μm² in the experiment). 100 points are randomly selected from the character area (QC) to simulate the distribution of SR, with each point occupying a circle with a radius of 3 or 4 pixels. The arrival times of photons are established for each pixel in the circular area with a specific repetition frequency. As demonstrated in Fig. 4a, photon counting is conducted for each timeline, with the results serving as the intensity values for each pixel to generate the raw data. These data are subsequently normalized to a range of 256. Subsequently, DFT processing was performed on each timeline in order to derive the modulation factor of the corresponding pixel, thus obtaining the frequency-domain imaging of the raw data, as illustrated in Fig. 4b. Next, incorporate the interference information (number of noise points: 1000) throughout the original image, as depicted in Fig. 4c. Notably, the photons of interference areas do not follow the modulation properties. Similarly, after DFT processing, the frequency-domain imaging is displayed in Fig. 4d. The simulation results exhibit a high degree of consistency with the experimental results, thereby validating the accuracy and reliability of the simulation process.

Secondly, a convolutional neural network (EfficientNet V2) was employed for the purpose of character recognition, with the specific objective of identifying characters containing different numbers of molecules [39]. EfficientNet V2 has been trained on a dataset consisting of 10,000 images (512×512 pixels) with different molecular distribution structures. The molecules are spatially arranged in accordance with the characters 0-9, A-Z, and any two combinations of A-Z in the dataset. The dataset is exemplified in Fig. 5, the original data and corresponding frequency-domain imaging of the character "QC" with varying numbers of molecules are depicted in Figs. 5a and 5b. Subsequently, the trained EfficientNet is applied to identify characters containing different numbers of molecules. The relationship between accuracy (Acc) and the average number of molecules is illustrated in Fig. 3f. When the average number of molecules <N> is 100, the recognition accuracy can reach 99.995%. Even if the number of molecules in a character is only 50, the accuracy of EfficientNet is still as high as 99.268%.

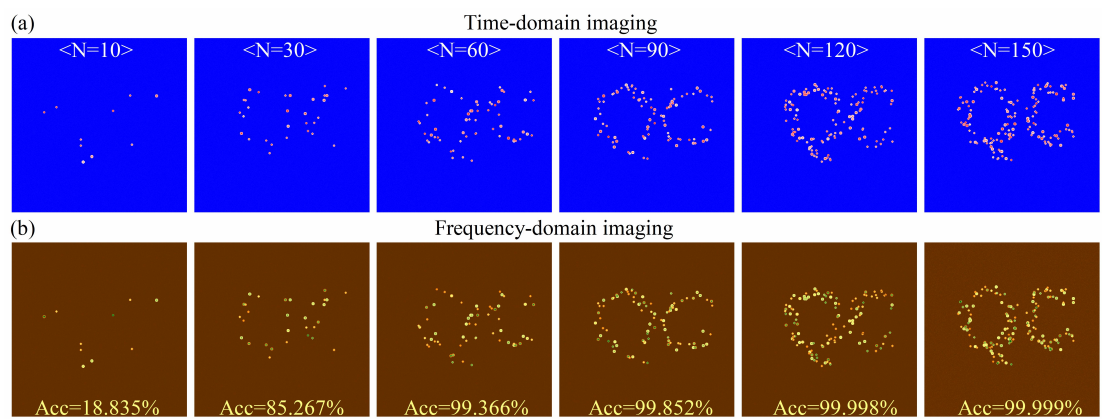

**Fig. 5**. **Comparison between time-domain and frequency-domain imaging under different average numbers of single molecules obtained by numerical simulation.** (a) Simulated time-domain imaging at the average molecular number of <N> = 10, 30, 60, 90, 120, 150, respectively; (b) the corresponding frequency-domain imaging and recognition accuracy.

As the number of molecules reduces, the accuracy decreases rapidly to unrecognizable. The decay process conforms to $e$ exponential distribution, as shown in Fig. 3f. The relationship between the *Acc* and the initial molecule number ($N_0$) can be expressed by the

following equation:

$$Acc(P,t) = \beta \cdot \exp\left[-N_0 \cdot e^{-\alpha \times p \times t} \cdot k_{dis}\right] + c \qquad (11),$$

where $p$ is the excitation power, $\beta$ is the attenuation coefficient, which is negative; $k_{dis}$ is the distortion rate; and $c$ is a constant, usually 1.

The number of molecules has a significant impact on the accuracy of information recognition, which is key to the "disposability" of anti-counterfeiting labels. Our results point out that we have developed a high-quality disposable micro-optical anti-counterfeiting technology based on SMQC. The technology overcomes the disadvantages of optical materials, such as broad spectral spread and susceptibility to environmental influences, and provides an effective anti-counterfeiting means to combat the rampant counterfeiting of products. On the other hand, to expand the potential applications of SMQC anti-counterfeiting labels, we employed a quenching inhibitor to suppress the photo-quenching of single molecules. As shown in Fig. 6a, we inserted glycerol-based mountant, PMMA, into the original sample structure to form a sandwich isolation layer. In this structure, the photo-quenching is substantially suppressed, and the reduction in the number of molecules during three consecutive acquisitions is negligible. Accordingly, in the presence of QDs interference, the recognition accuracy of the labels is consistently above 99% (as shown in Fig. 6b), *i.e.,* the transition from disposable to reusable is realized.

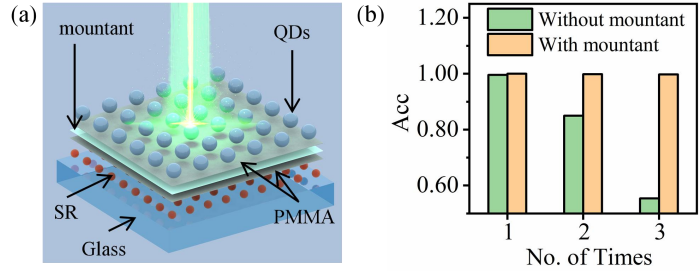

**Fig. 6. Reusable labels.** (a) Structure illustration for anti-counterfeiting labels. Substrate: glass; first layer: SR molecules; second layer: polymethylmethacrylate (PMMA); third layer: glycerol-based mountant; fourth layer: PMMA; top layer: QDs. (b) In comparison to the absence of a quenching inhibitor, the decrease in the number of molecules with the acquisition times can be neglected, and the recognition accuracy is consistently above 99%.

## 6. Conclusion:

In this study, we proposed a new method to fabricate micro-optical anti-counterfeiting labels based on SMQC. SMQC-based imaging offers high contrast and robust resistance to background interference. Single-molecule labels are prepared by inkjet printing technology to achieve molecular-level information writing. In order to conceal the real information, QDs are selected as interference due to their complex dynamics. We use SMQC as the carrier of anti-counterfeiting information, then read the information by the DFT transformation of the fluorescence photon arrival time of the molecule. The experimental results demonstrate that micro-optical labels based on SMQC have excellent anti-counterfeiting capabilities and high levels of security. Subsequently, we reproduced the experimental results using numerical simulations. The results show that even when the number of molecules is as low as 50, the level of recognition accuracy remains above 99%. Our research shows that SMQC-based labels are a proven disposable micro-optical anti-counterfeiting technology that is designed to combat counterfeiting and can enhance the security of expensive products such as new pharmaceuticals, highly sophisticated electronic devices, and luxury goods. In addition to the study of disposable labels, we also attempted to employ glycerol-based mounts during sample preparation to inhibit molecular quenching due to laser irradiation. This allowed us to extend the applicability of the scheme to non-disposable scenarios. The advancement of intelligent manufacturing has led to the refinement of atom and molecule manipulation techniques, including optical tweezers, magnetic tweezers, and scanning tunneling microscopes. This has paved the way for the meticulous construction of single-molecule arrays. By integrating emerging molecular manipulation techniques, we expect to achieve more precise information encoding, provide a wider range of information dimensions, and facilitate the development of more sophisticated anti-counterfeiting techniques.

**Appendix**

*AppendixA: SMQC*

Dye molecules can be treated as a two-level system [23], consisting of a ground state and an excited state, as shown in Figure 1a. Compared to other materials, a single molecule is considered an ideal system for investigating ultrafast physics due to its relatively simple energy diagram. As shown in Figure 1b, different points on the Bloch sphere can be used to represent various coherent superposition states of the molecule. The two poles correspond to two eigenstates ("South Pole", ground state, $|0\rangle$; "North Pole", excited state, $|1\rangle$). The remaining points on the sphere signify the coherent superposition states of the electron wave functions between the ground and excited states [41, 42]. The interaction between the laser pulse and a single molecule corresponds to the rotation of the Bloch vector from the ground state around the center of the circle [41], as shown in the arc arrow in Figure 1a. Initially, the single molecule populates in the ground state, and its initial wave function can be expressed as:

$$|\psi_1\rangle = \begin{pmatrix} 0 \\ 1 \end{pmatrix} \qquad (12).$$

Neglecting the effects of dephasing, the electron population probability $\rho_{ee}$ of the excited state can be controlled by the relative phase $\Delta\varphi$. The final expression is [27, 42]:

$$\rho_{ee} = \frac{1}{2} \cdot \left(1 - \cos\theta_1 \cos\theta_2 + \sin\theta_1 \sin\theta_2 \cos\Delta\varphi\right) \tag{13}$$

$\theta_1$ and $\theta_2$ represent the pulse regions of the interaction between two pulses and a single molecule, respectively[42]; $\Delta\varphi$ is the phase difference between two pulses.

In practice, however, individual molecules tend to interact with their surroundings, leading to the loss of phase information, *i.e.*, the process of dephasing. To quantify the ability to produce coherent superposition states in the presence of dephasing, we introduce the quantum coherence visibility $V$. When the single molecule fully maintains the quantum coherence properties, $V$ is 1; and when the quantum coherence information of the single molecule is completely lost, $V$ is 0. In accordance with the principles of quantum coherence theory and the extant literature on the subject [41, 42, 43], when the pulse width is considerably smaller than the time scale of single-molecule dynamics, the quantum coherence visibility can be expressed as follows: $V = e^{-\Delta t/T_2^*}$, where $T_2^*$ is the dephasing time. It is evident that quantum coherent visibility is directly influenced by the dephasing time. In the absence of dephasing and under the assumption that both pulses possess an equal degree of interaction with a single molecule, *i.e.*, when $\theta_1 = \theta_2 = \theta$, the equation (2) can be simplified as follows:

$$\rho_{ee} = \frac{1}{2} \cdot \sin^2\theta \cdot \left(1 + \cos\Delta\varphi\right) \tag{14}$$

When the dephasing is taken into account, the equation can be modified as:

$$\rho_{ee} = \frac{1}{2} \cdot \sin^2\theta \cdot \left(1 + V \cdot \cos\Delta\varphi\right) \tag{15}$$

Considering the PL intensity of molecules is proportional to electron population probability $\rho_{ee}$, the quantum coherence properties can be explored by performing their PL intensity. However, the PL intensity of molecules is usually influenced by many factors, including chemical and physical stability (photon blinking) [26], transition dipole moment [27], and energy transfer within the surrounding environment [28]. The periodic modulation of the relative phase $\Delta\varphi$ effectively eliminates these effects, and the quantum coherence information in the PL can be rapidly obtained by DFT, as illustrated in Fig. 1c. The modulation intensity $f(\omega)$ at frequency $\omega/2\pi$ can be obtained by performing DFT on the arrival time of each photon:

$$f(\omega) = \left|\sum_{n=1}^{N} e^{-i \cdot t_n \cdot \omega}\right| \tag{16}$$

Where $N$ is the total number of photons. When $\omega t \gg 2\pi$, the relationship between the single-molecule modulation intensity and the total photon number and quantum coherence visibility is:

$$f(\omega) = \frac{1}{2} V \cdot N \tag{17}$$

Through this method, quantum coherence information and the corresponding frequency domain imaging can be quickly obtained. As illustrated in Fig. 1d, the respective frequency spectrum of the dye molecule (squaraine-derived rotaxane (SR)) and the background were obtained through modulation measurements. A particularly straightforward outcome is that at the modulation frequency (1 kHz), the modulation intensity of the single molecule is markedly greater than that of the background. Figure 1e shows the wide-field PL imaging of SR prepared by spin coating, while Fig. 1f depicts the frequency-domain imaging calculated by DFT. It can be found that the imaging contrast in the frequency domain is significantly improved compared with time-domain imaging, which is consistent with previous reports [29, 30]. Our recent works show that this method can achieve more than 300-fold improvement in imaging contrast under the premise of strong background, showing great potential in suppressing background noise [27, 31].

*AppendixB: Implementation of anti-counterfeiting mechanism*

From the analysis of SMQC, it is clear that coherence visibility is crucial for distinguishing the signal from the background. Based on this, we propose an anti-counterfeiting method that utilizes samples with varying coherence visibility to achieve time-domain concealment of target information and precise extraction of information in the frequency domain. Specifically, inkjet printing technology is used to arrange single molecules according to the spatial distribution of the target information, and a second layer with lower coherence visibility and a disordered distribution is applied to mask the target information. Due to the difference in coherence visibility between the two layers, the concealed information in the time domain can be successfully extracted from the frequency-domain image. In the following, we will use single molecules and quantum dots as examples to analyze the causes of their coherence visibility differences and theoretically validate the feasibility of this anti-counterfeiting approach.

The dipole moment is one of the primary driving forces behind dephasing, as its presence facilitates the coupling of a quantum system to the surrounding environment [32]. The strength of this coupling is determined by both the magnitude and orientation of the dipole moment. The core process involves the interaction between the system and the environment, leading to the loss of coherence in the quantum state. In dipolar systems, this coupling mainly occurs between the system's dipole moment $\overline{P}$ and environmental factors such as electric fields, phonons, and other interactions. The system's Hamiltonian can be written as:

$$\overline{H} = \overline{H}_{sys.} + \overline{H}_{env.} + \overline{H}_{int.} \tag{18}$$

where $\overline{H}_{sys.}$ describes the free evolution of the quantum emitters, while $\overline{H}_{env.}$ represents the free evolution of the environment (such as phonons or electric fields). $\overline{H}_{int.}$ describes the coupling between the system and the environment. The coupling term $\overline{H}_{int.}$ is given by: $\overline{H}_{int.} = -\overline{P} \cdot \overline{E}_{env.}$, Where $\overline{E}_{env.}$ is the local electric field in the environment. The noise in the environment introduces random fluctuations that lead to dephasing in the quantum system.

Assuming that the ambient noise acts on the system dipole moment through random electric field fluctuations $\vec{E}(t)$, the autocorrelation function of the environment is [33, 34]:

$$\langle \vec{E}_i(t)\vec{E}_i(t+\tau)\rangle = \int J_i(\omega)e^{i\omega\tau}d\omega \tag{19}$$

Where $\tau$ is the time delay, $J_i(\omega)$ is the noise spectral density, describing the coupling strength of the environment in direction i. $T_2^*$ is determined by the value of $J_i(\omega)$ at zero frequency.

The direction of the dipole moment of a single molecule is primarily determined by the distribution of electrons within the molecule. This is attributable to the rigid structure of the molecule, which results in a dipole moment in space that typically exhibits a single definitive direction, assuming that along the z-direction, then [34, 35]:

$$\frac{1}{T_2^{SM}} \propto J_z(0) \tag{20}$$

If the ambient noise is white noise independent in all directions with intensity $S_{SM}$, then the dephasing time of single molecules can be written as:

$$T_2^{SM} = J_z(\omega) = \frac{1}{S_{SM}} \tag{21}$$

QDs are typically quasi-spherical, ellipsoidal, or possess certain lattice symmetries, allowing the dipole moment to exist independently along three spatial directions. This structural property introduces multi-channel environmental noise effects, leading to an accelerated dephase rate. Within the framework of quantum mechanics, the dipole moment can be expressed in terms of the wavefunctions of electrons and holes [36, 38]:

$$\vec{P} = -e\int \psi_e^*(\vec{r})r\psi_h(\vec{r})d^3\vec{r} \tag{22}$$

where $\psi_e$ and $\psi_h$ are the wave functions of electrons and holes. The coupling strength between the system and the environment is:

$$H_{interaction} = -\vec{P}\cdot\vec{E}_{env} = -\left(P_x\hat{E}_x + P_y\hat{E}_y + P_z\hat{E}_z\right) \tag{23}$$

Given the independence of the contributions from each direction, the total dephasing rate is a superposition of the three directions:

$$\frac{1}{T_2^{QD}} = \frac{1}{T_{2,x}} + \frac{1}{T_{2,y}} + \frac{1}{T_{2,z}} \propto J_x(0) + J_y(0) + J_z(0) \tag{24}$$

Where $T_{2,x}$, $T_{2,y}$, $T_{2,z}$ are the dephasing times in the x, y, and z directions, respectively. If the ambient noise is white noise independent in all directions, i.e., $J_i(\omega) = S_{QD}$ ($i = x, y, z$), then the total dephasing rate and time can be expressed as:

$$\frac{1}{T_2^{QD}} = 3S_{QD} \tag{25}$$

$$T_2^{QD} = \frac{1}{3S_{QD}} \qquad (26).$$

Moreover, the PL of QDs is relatively complex, involving multiple physical processes such as bright and dark excitons, biexcitons, and gray states [36, 37]. These processes cause the coherent information of QDs to become obscured by other incoherent behaviors. In summary, QDs exhibit short dephasing times, typically in the picosecond range [34, 38]. In contrast, single molecules have significantly longer dephasing times, generally in the nanosecond range.

The dephasing time $T_2^*$ characterizes the timescale over which a quantum system loses coherence, reflecting the rate at which environmental interactions induce phase randomization. During dephasing, quantum coherence visibility decays exponentially with time. A shorter $T_2^*$ indicates faster environment-induced coherence loss, leading to a rapid decline in coherence visibility and making it challenging to preserve quantum properties. Conversely, a longer $T_2^*$ allows the system to maintain coherence for a more extended period, enhancing the observability of coherence visibility in experiments. This property enables efficient anti-counterfeiting applications through careful material selection. In this paper, we selected SR molecules as anti-counterfeiting information carriers and CdTeSe/ZnS QDs (CdTeSe/ZnS QDs@800nm, 800QDs) as interference information. The contrast of frequency-domain imaging is adjusted by using the natural difference between the two in dephasing time.

*AppendixC: EfficientNet V2*

EfficientNet V2 is a highly efficient convolutional neural network architecture designed to achieve high accuracy while maintaining computational efficiency. It is particularly suitable for tasks requiring real-time processing, such as image classification, object detection, and segmentation. Compared to its predecessor, EfficientNet V1, EfficientNet V2 introduces new modules (e.g., FusedMBConv) and optimized strategies (e.g., progressive learning), significantly improving training speed and inference performance. These advancements make it well-suited for large-scale data processing and resource-constrained environments. The process is shown in Fig. A-1, and the detail about Figs A-1 a-c are described below:

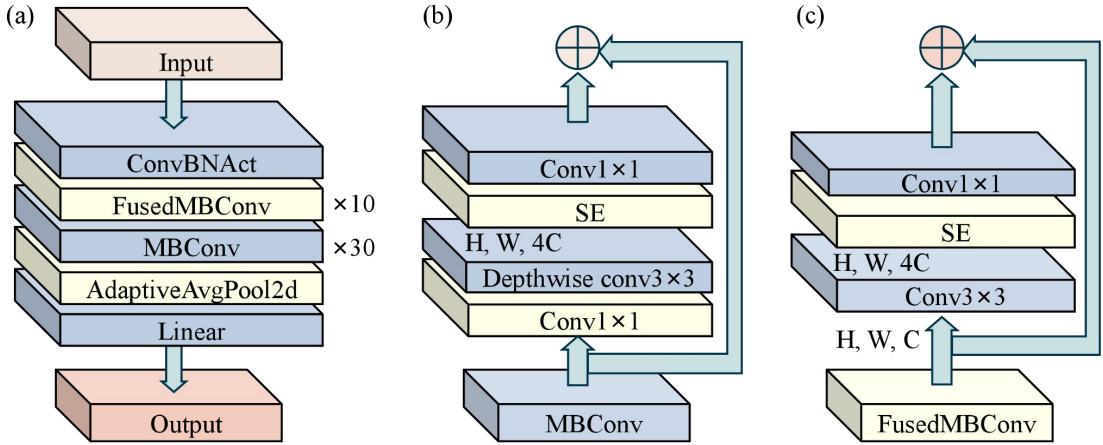

Fig. A-1. Schematic diagram of EfficientNet V2. (a) Overall Network Architecture. (b) MBConv

Module Mechanism. (c) FusedMBConv Module Mechanism.

**Overall Network Architecture**

1. Input

    The input image, after preprocessing, is fed into the network.

2. ConvBNAct

    This module combines convolution (Conv), batch normalization (BN), and activation (e.g., ReLU) layers. It extracts low-level features while ensuring the stability of data distribution.

3. FusedMBConv (×10)

    A key module introduced in EfficientNet V2, FusedMBConv replaces depthwise separable convolution with standard convolution. This design improves runtime efficiency, especially for small models and high-resolution tasks.

4. MBConv (×30)

    The MBConv module (detailed in Figure b) employs depthwise separable convolution to extract features efficiently, with the addition of a Squeeze-and-Excitation (SE) block to enhance feature representation.

5. AdaptiveAvgPool2d and Linear Layers

    The global average pooling layer reduces the feature map to a fixed-size vector, followed by a linear layer for classification.

**MBConv Module Mechanism**

1. 1×1 Convolution

    Performs channel-wise dimensionality reduction or expansion, enabling flexible feature representation.

2. Squeeze-and-Excitation (SE) Block

    Dynamically adjusts the importance of feature channels by leveraging global context information.

3. Depthwise Convolution (3×3)

    Performs convolution independently on each channel, significantly reducing computational costs.

4. Skip Connection

    Facilitates gradient flow and enhances the integration of low-level and high-level features, ensuring stability during training.

**FusedMBConv Module Mechanism**

1. 1×1 Convolution

    Similar to MBConv, this step adjusts the number of channels.

2. Standard Convolution (3×3)

    Replaces depthwise separable convolution to improve parallel computation efficiency, particularly in scenarios requiring low latency or smaller models.

3. SE Block and Skip Connection

    The SE block and skip connection are retained to ensure robust feature extraction and gradient stability.

**Funding:** National Natural Science Foundation of China (Nos. U23A20380, U22A2091, 62222509, 62005150, 62127817, and 6191101445), National Key Research and Development

Program of China (Grant No. 2022YFA1404201), Shanxi Province Science and Technology Innovation Talent Team (No. 202204051001014), 111 projects (Grant No. D18001) and Shanxi Provincial Basic Research Program Project (202203021222107).

**Data availability:** Data underlying the results presented in this paper are not publicly available at this time but may be obtained from the authors upon reasonable request.

**Disclosures:** The authors declare no conflict of interest.